\documentclass[
twocolumn,
secnumarabic,
amssymb,
amsmath,
bm,
natbib, 
nobibnotes, 
aps, 
prx,
10pt,
superscriptaddress]{revtex4-2}

\usepackage{graphicx} % Required for inserting images
\PassOptionsToPackage{hyphens}{url}\usepackage{hyperref}
\usepackage{upgreek}

\usepackage{siunitx}
\DeclareSIUnit\bar{bar}

\usepackage{amsmath}
\usepackage{siunitx}
\usepackage{lineno}
\usepackage{makecell}

\DeclareSIUnit\bar{bar}

\begin{document}
\newcommand{\red}[1]{\textcolor{red}{#1}}
\newcommand{\green}[1]{\textcolor{green}{#1}}
\newcommand{\blue}[1]{\textcolor{blue}{#1}}
\newcommand{\maxence}[1]{\textcolor{blue}{[MT: #1]}}
\newcommand{\mathis}[1]{\textcolor{red}{[MM: #1]}}

% \title{Hydrodynamic simulations of discharge plasma capillaries}
%\title{Simulations of energy flows in discharge capillaries for plasma accelerators}
\title{Characterization of discharge capillaries via benchmarked hydrodynamic plasma simulations}

\author{S. M. Mewes} %ORCID: 0000-0002-9625-7382
\email{mathis.mewesmdi@desy.de}
\affiliation{Deutsches Elektronen-Synchrotron DESY, Notkestraße 85, 22607 Hamburg, Germany}
\affiliation{Institut für Experimentalphysik, Universität Hamburg, Luruper Chaussee, 22761 Hamburg, Germany}

\author{G. J. Boyle} %ORCID: 0000-0002-8581-4307
\affiliation{James Cook University, 1 James Cook Dr, Townsville City QLD 4814, Australia}

\author{R. D'Arcy} %ORCID: 0000-0002-7224-8334
\thanks{Now at University of Oxford, Department of Physics, Keble Road, Oxford OX1 3RH, United Kingdom}
\affiliation{Deutsches Elektronen-Synchrotron DESY, Notkestraße 85, 22607 Hamburg, Germany}

\author{J. M. Garland} %ORCID: ?
\affiliation{Deutsches Elektronen-Synchrotron DESY, Notkestraße 85, 22607 Hamburg, Germany}

\author{M. Huck} %ORCID: ?
\affiliation{Deutsches Elektronen-Synchrotron DESY, Notkestraße 85, 22607 Hamburg, Germany}

\author{H. Jones} %ORCID: 0000-0003-4480-3449
\affiliation{Deutsches Elektronen-Synchrotron DESY, Notkestraße 85, 22607 Hamburg, Germany}

\author{G. Loisch} %ORCID: ??? 0000-0002-6178-2339 ???
\affiliation{Deutsches Elektronen-Synchrotron DESY, Notkestraße 85, 22607 Hamburg, Germany}

\author{A. R. Maier} %ORCID: 0000-0003-3361-4247
\affiliation{Deutsches Elektronen-Synchrotron DESY, Notkestraße 85, 22607 Hamburg, Germany}

\author{J. Osterhoff} %ORCID: 0000-0002-7684-0140
\thanks{Now at Lawrence Berkeley National Laboratory, 1 Cyclotron Rd, Berkeley, CA 94720, USA}
\affiliation{Deutsches Elektronen-Synchrotron DESY, Notkestraße 85, 22607 Hamburg, Germany}

\author{T. Parikh} %ORCID: ?
\affiliation{Deutsches Elektronen-Synchrotron DESY, Notkestraße 85, 22607 Hamburg, Germany}

\author{S. Wesch} %ORCID: 0000-0002-3789-8884
\affiliation{Deutsches Elektronen-Synchrotron DESY, Notkestraße 85, 22607 Hamburg, Germany}

\author{J. C. Wood} %ORCID: 0000-0003-4413-7044
\affiliation{Deutsches Elektronen-Synchrotron DESY, Notkestraße 85, 22607 Hamburg, Germany}

\author{M. Thévenet} %ORCID: 0000-0001-7216-2277
\affiliation{Deutsches Elektronen-Synchrotron DESY, Notkestraße 85, 22607 Hamburg, Germany}

\date{\today}

\begin{abstract}
Plasma accelerators utilize strong electric fields in plasma waves to accelerate charged particles, making them a compact alternative to radiofrequency technologies. Discharge capillaries are plasma sources used in plasma accelerator research to provide acceleration targets, or as plasma lenses to capture or focus accelerated beams. They have applications for beam-driven and laser-driven plasma accelerators and can sustain high repetition rates for extended periods of time. Despite these advantages, high-fidelity simulations of discharge capillaries remain challenging due to the range of mechanisms involved and the difficulty to diagnose them in experiments. In this work, we utilize hydrodynamic plasma simulations to examine the discharge process of a plasma cell and discuss implications for future accelerator systems. The simulation model is validated with experimental measurements in a 50-mm-long, 1-mm-wide plasma capillary operating a 12-27\,kV discharge at 2-12\,mbar hydrogen pressure. For 20\,kV at 8.7\,mbar the discharge is shown to deposit 178\,mJ of energy in the plasma. Potential difficulties with the common density measurement method using $\text{H}_\alpha$ emission spectroscopy are discussed. This simulation model enables investigations of repeatability, heat flow management and fine tailoring of the plasma profile with discharges.
\end{abstract}

% Discharge capillaries are a common device used to generate plasmas. While their role in plasma accelerators is well established, advanced plasma shaping, e.g. for active plasma lenses (APL) or injection schemes, requires improved understanding and modelling of the processes at play.

\maketitle
\section{Introduction}
\label{sec:intro}
%\begin{itemize}
%    \item Main point: HD simulations help understanding and developing discharge plasma sources for applications in future accelerators. (HALHF link? high average power?)
%    \item Previous hydrodynamic models make some mistakes and lack experimental benchmarking.
%    \item Necessary enhancements (Boundary conditions, Reaction system, ...)
%    \item OES measurement of electron density is inaccurate
%\end{itemize}

Plasma accelerators\,\cite{Tajima:1979,Chen:1985} are compact sources of relativistic electron beams, where the accelerating gradients can be orders of magnitude higher than rf-based technologies. Proof-of-principle experiments and design studies demonstrated the potential of electron beams accelerated by laser-driven (LPAs) or beam-driven plasma accelerators\,\cite{Blumenfeld:2007} for applications such as high-energy photon sources\,\cite{Corde:2013}, quantum electrodynamics\,\cite{Vranic:2014}, free-electron lasers\,\cite{Wang:2021,Pompili:2022}, and high-energy physics\,\cite{Schroeder:2010,Foster:2023}.
Progress in plasma accelerator performance has been strongly coupled to advances in driver technology\,\cite{Maier:2020,Lindstrom2025}, but recent efforts demonstrated the importance of controlling the plasma source\,\cite{Gonsalves:2016,Lindstrom:2024}. In particular, plans for plasma-based particle colliders\,\cite{Foster:2025}, free-electron lasers\,\cite{Galletti:2024,Assmann:2020}, and plasma injectors for storage rings\,\cite{Antipov:2021,FerranPousa:2022,MartinezDeLaOssa:2025,Winkler:2025} highlighted the importance of accurate and stable tailoring of the plasma source. Furthermore, bridging the gap between test accelerators and application-ready facilities requires high-repetition-rate operation, and efforts in this direction are being pursued by the community to explore feasibility and examine the constraints on the plasma source\,\cite{Rovige:2020,Jakobsson:2021,DArcy:2022}.

Beside shaping the gas profile with advanced gas cell or gas jet designs, two methods are commonly used for the tailoring of the plasma source prior to the wakefield acceleration event. In the first, a carefully shaped laser pulse ionizes a thin filament of plasma within a gas volume, which can be used to create a guiding structure\,\cite{Durfee:1993,Shalloo:2018}. In the second, a high-voltage discharge initiates an electrical breakdown in a long and thin capillary\,\cite{Leemans:2006,Butler:2002}. This discharge technology remains under consideration for plasma acceleration due to its favorable scalings to long plasmas and high repetition rates\,\cite{Gonsalves:2016}, as well as its versatility for plasma lenses\,\cite{VanTilborg:2015} and electrothermal plasma jets\,\cite{Winfrey:2012}. 

Simulations of these devices typically rely on magneto-hydrodynamics (MHD) or reduced models\,\cite{Bobrova:2001,Cook:2020,Boyle:2021,Bagdasarov:2017}, and are an active part of plasma acceleration research\,\cite{Gonsalves:2019}. However, systematic benchmarking of simulations against experiments, including all physical processes, remains uncommon in the literature\,\cite{Arjmand:2025} due to experimental and multi-physics simulation challenges, limiting the ability to provide comparable results. For instance, investigations of demanding regimes about fine plasma shaping and heat management, as required for HALHF\,\cite{Foster:2023}, remain extremely difficult.

In a previous work, we proposed the HYQUP (HYdrodynamic QUasineutral Plasma) model\,\cite{Mewes:2023}, a collisional plasma simulation framework implemented in the 1d/2d/3d/cylindrical COMSOL Multiphysics$^\text{®}$ software\,\cite{COMSOL}. Using finite elements on COMSOL's advanced meshing capabilities, the model solves for plasma fluid equations in a single-fluid, two-temperature (electrons and \textit{heavies}, i.e. all neutrals and ions) model. Unlike \emph{magneto-}hydrodynamics, HYQUP assumes that the dynamics is not determined by the magnetic field but rather pressure/temperature effects, easing the requirements on the solvers. Electric and magnetic corrections can be added separately. This regime is appropriate for hydrodynamic optical-field-ionized (HOFI) waveguides studied in Ref.\,\cite{Mewes:2023} as well as the discharge regime of the present work (systems where the dynamics is governed by the magnetic field, like Z-pinch\,\cite{Haines:2011}, may require a different approach). HYQUP also models the reaction statistics (for ionization, recombination, dissociation, etc.) rather than assuming a Saha equilibrium, allowing for simulating extreme processes.

In this work, we present a simulation method comprising the electric discharge, the plasma response and the emission of hydrogen spectral lines based on HYQUP, where the relevant processes (discharge and interaction with the wall) were added. Through extensive benchmarks with experimental results, we demonstrate the importance of accurate boundary conditions and spectral line calculations to reproduce measurements, and show that transverse spectroscopy measurements may be imprecise due to plasma inhomogeneity. This study also clarifies the flow of energy in the plasma during each discharge event in conditions relevant for plasma acceleration.

Section\,\ref{sec:model} presents the additions to the HYQUP model. The experimental setup and simulation pipeline used for the benchmark are detailed in Sec.\,\ref{sec:bench}. Considerations on the discharge physics and its consequences are explored in Sec.\,\ref{sec:phys}.
% In this work, we present a simulation method comprising the electric discharge, the plasma response and the emission of hydrogen spectral lines. Through extensive benchmarks with experimental results, we demonstrate the importance of accurate boundary conditions and spectral line calculations to reproduce measurements. The simulation, based on the HYQUP (HYdrodynamic QUasineutral Plasma) model\,\cite{Mewes:2023} for nanosecond to millisecond timescale dynamics in collisional, near atmospheric plasmas, implemented in the COMSOL Multiphysics$^\text{®}$ software\,\cite{COMSOL}, is used to clarify the dominating mechanisms and energy transfers in the process. The HYQUP simulation model is extended as outlined in in Sec.\,\ref{sec:model}, to include important mechanisms of the discharge. The experimental setup and simulation pipeline used for the benchmark are detailed in Sec.\,\ref{sec:bench}. Considerations on the discharge physics and its consequences are explored in Sec.\,\ref{sec:phys}.

\section{Simulation Model}
\label{sec:model}

%\begin{itemize}
%    \item \green{Cell operation cycle (No wake)}
%    \item \green{Only Hydrogen, but similar for others.}
%    \item Generic cell geometry and surrounding conditions (Vacuum, electric insulator, high thermal conduction, inlets for gas delivery)
%    \item \green{Discharge stages: Ionization, steady state, Recombination}
%    \item \green{longitudinal dynamics: expulsion, asymmetry of current}
%    \item Applicability window (Low ionization: Neutrals contribute, initial state dependency, no steady state. High ionization: Plasma-Wall interaction not done properly. Low density: non-collisional plasma)
%    \item Low Ionization: Neutral contributions in mixture.
%    \item \textbf{Model}
%    \item \green{Expansion of the HYQUP model} 
%    \item \green{Electrical Current}
%    \item \green{Ohmic Heating}
%    \item \green{Current self focusing}
%    \item \green{Magnetization}
%    \item \green{New reaction model: NRL ionization, Molecular recombination}
%    \item \green{General setup (Geometry, Symmetries)}
%    \item \green{Floating boundary sheath model}, 
%    \item \green{Fig 5a): OES NOT a reliable measure of center density (important for PWA)}
%\end{itemize}

\begin{figure}[b]
    \centering
    \includegraphics[width=\columnwidth]{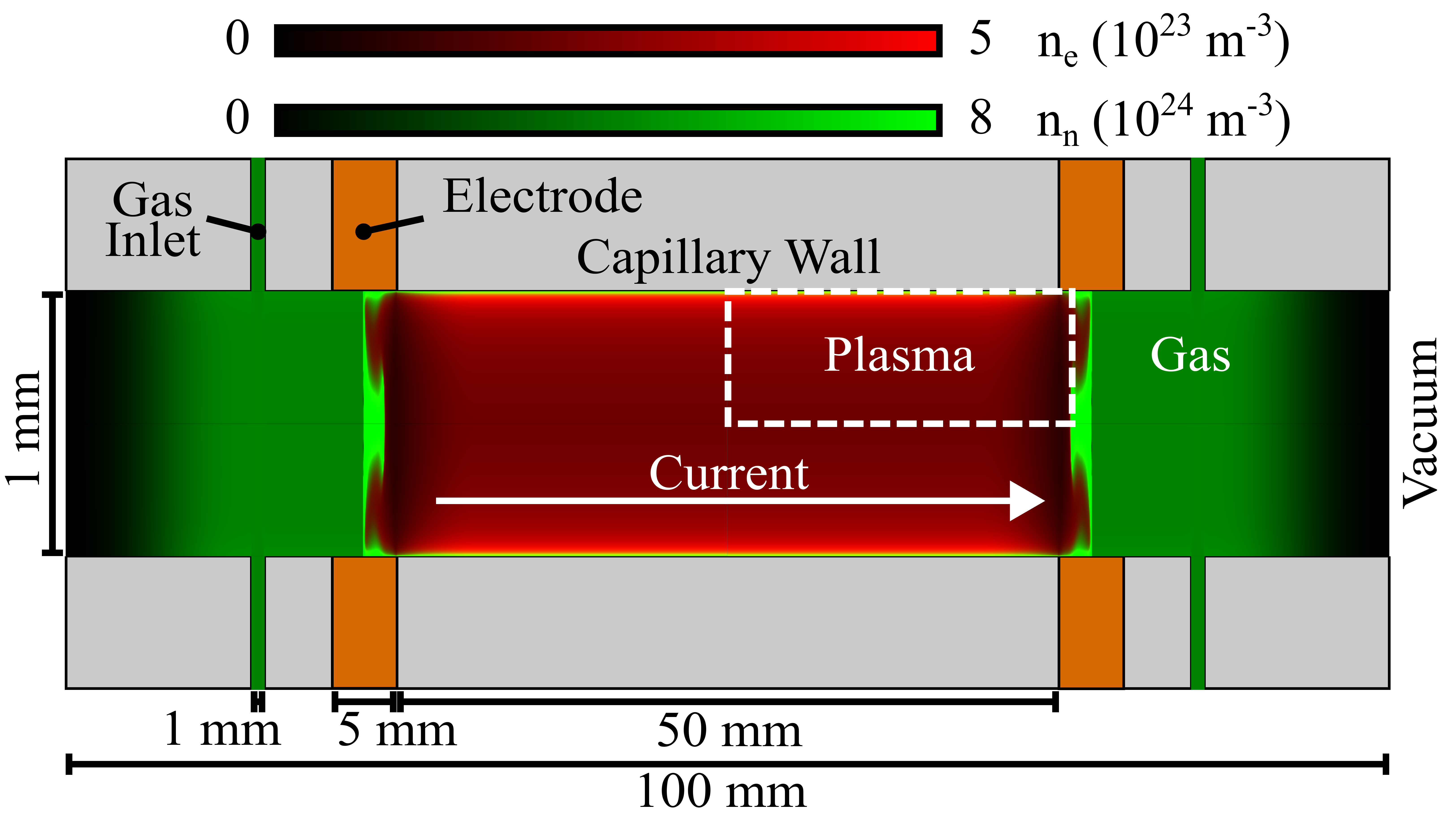}
    \caption{Schematic of the discharge capillary. The inset simulation example shows the electron density $n_e$ and neutral atom density $n_n$ near the peak of a discharge (after 350\,ns). The dotted line represents the simulation box used for the benchmark, assuming cylindrical symmetry.}
    \label{fig:cell}
\end{figure}
%Needs electrode width and current direction.

% What is a discharge source?
A discharge capillary, like the one shown in Fig.\,\ref{fig:cell}, is a long, thin channel made of an electrical insulating material such as sapphire. It is filled with gas supplied by the inlet channels and a high voltage is applied between the electrodes to trigger an electrical breakdown. The gas dynamics involved in filling the cell occur on µs-ms timescales, while the plasma fluid dynamics of the discharge takes place on a shorter ns-µs timescale, which is the focus of this article. A third regime, the kinetic dynamics of a wakefield acceleration event occurring on an even shorter (fs-ps) timescale, is not considered here.

% Summary of HYQUP and new additions
The HYQUP model was updated in order to make it applicable to capillary discharges. The main enhancements, detailed below, are the introduction of an electric current flowing through the plasma, the confinement of the plasma inside a thin capillary, and improvements to the reaction system.

% Implementation of the discharge current
\subsection{Electric current}
The discharge current is primarily carried by electrons, which move faster than ions due to their lower mass. The ion current is therefore neglected. 
%In the simulation model the electrons are modeled implicitly through the assumption of quasi-neutrality and the effects of the current are added as external influences. 
The current density $\Vec{J}$ and electric field $\Vec{E}$ are computed using Ohm's law $\vec{J} = \sigma_{e} \vec{E}$ and Gauss' law $\vec{\nabla}\cdot\vec{E} = 0$,
where $\sigma_e$ is the electron electrical conductivity. A specified total current $I(t)$ flowing through the electrode surfaces is set as a boundary condition
\begin{equation}
    \int_{\text{Cathode}}\vec J \cdot \vec{n} \, dA = -\int_{\text{Anode}}\vec J \cdot \vec{n} \, dA \equiv I(t),
    \label{eq:electrode_current}
\end{equation}
where $\vec n$ is the normal vector on the surface element $dA$.

The electrical conductivity, like other transport parameters in the original HYQUP model (local dynamic properties like thermal conductivity, diffusivity or viscosity), is calculated by statistical consideration of particle collisions in the plasma. The total resistivity is estimated by summation of the resistivities $(\sigma_{e\alpha})^{-1}$ between electrons and all other particle species $\alpha$. The total electrical conductivity is then
\begin{align}
    \sigma_e &= \left( \sum_\alpha (\sigma_{e\alpha})^{-1} \right)^{-1} \\
    \sigma_{e\alpha} &= \gamma_{\text{el}, e\alpha}\frac{n_e e^2}{m_{e\alpha}} \frac{1}{\nu_{e\alpha}},
    \label{eq:electrical_conductivity}
\end{align}
where $\gamma_{\text{el}, e\alpha}$ is a prefactor for additional effects (e.g. magnetization), $n_e$ is the number density of electrons, $e$ is the elemental charge, $m_{e\alpha}=\frac{m_e m_\alpha}{m_e + m_\alpha}$ is the reduced mass, and $\nu_{e\alpha}$ is the momentum transfer collision rate between electrons and species $\alpha$, see Ref.\cite{Mewes:2023} for details.

% Ohmic Heating
Ohmic heating is the dominant effect by which the current affects the plasma. The resistive energy losses of the electron current are converted into electron heat, which drives the ionization and flow of the plasma. The Ohmic heating is calculated as
\begin{equation}
    Q_{\Omega} = \vec{J}\cdot\vec{E} = \frac{1}{\sigma_{e}} J^2
    \label{eq:OhmicHeat}
\end{equation}
and included as a source term in the electron heat equation.

% Magnetic field
\subsection{Magnetic field}
The electric current influences the plasma through the magnetic field in two ways: The first is the Lorentz force on the current itself, the second is \emph{magnetization}, which affects the transport properties. Both effects are generally weak within the parameter space observed in the benchmark below, but may be non-negligible at specific times and locations. These effects are more important in high-current-density applications, such as active plasma lenses.

% Self-focussing
The magnetic Lorentz force acts on the electron current, transversely compressing the quasi-neutral plasma. The force density on the fluid is implemented as $\vec{f_\text{L}} = \vec{J} \times \vec{B}$. In high-current discharges, where the magnetic pressure is of similar or higher magnitude to the thermal pressure, the self-focusing of the electron current leads to pinching.

% Magnetization
The magnetization of the plasma happens due to the individual charged particles spiraling around magnetic field lines. This constrains the particles motion range and collision rates and, consequently, alters the macroscopic transport parameters. 
The magnitude of the effects depends on the Hall parameter $\beta_{\text{Hall}} = d_{\text{mfp}}/r_{\text{mag}}$, with the mean free path $d_{\text{mfp}}$ and Larmor radius $r_{\text{mag}}$. For the slow and heavy ions, the Hall parameter is small ($\beta_{\text{Hall}}\ll1$) such that magnetization is negligible. But, for electrons, it can be non-negligible (up to 0.1) in the early discharge and close to the plasma boundaries. For high-current applications, where $\beta_{\text{Hall}} > 1$ is common, the magnetization becomes a major factor.
This effect is implemented, following suggestion by Diaw et al.\,\cite{Diaw2022}, by including Eq.\,A2 from Davies et al.\,\cite{Davies2021} in the $\gamma_{\text{el}, e\alpha}$ prefactors of the electrical conductivity and similarly Eq.\,83e from Ji et al.\,\cite{Ji2013} in the thermal conductivity. The anisotropy of these models is neglected in the cylindrically symmetric capillary geometry, applying the component orthogonal to the magnetic field as a scalar.

% Reaction System enhancement
\subsection{Reactions}
\renewcommand{\arraystretch}{2.25}
\begin{table}[tb]
    \centering
    \begin{tabular}{c|c}
        Reaction &\hspace{5px} Source  \\
        \hline\hline
        H$^+$ + 2e$^-$ $\rightarrow$ H + e$^-$ & 
        \makecell{Zel’dovich \& Raizer\,\cite{zel2002},\\(p.55, Eq.15 in NRL\,\cite{NRL2019}) }\\%\hline
        
        H + e$^-$ $\rightarrow$ H$^+$ + 2e$^-$ & Detailed Balancing\\\hline
        
        H$^+$ + e$^-$ $\rightarrow$ H + $\gamma$ & 
        \makecell{McWirther\,\cite{McWhirter:InHuddlestone1965}, \\(p.55, Eq.14 in NRL\,\cite{NRL2019})}\\\hline
        
        H$_2$ + e$^-$ $\rightarrow$ 2H + e$^-$ & 
        \makecell{McWirther\,\cite{McWhirter:InHuddlestone1965}, \\(p.55, Eq.12 in NRL\,\cite{NRL2019})}\\%\hline
        
        2H + e$^-$ $\rightarrow$ H$_2$ + e$^-$ & 
        Detailed Balancing\\\hline
        
        H$_2$ + H $\rightarrow$ 2H + H & 
        \makecell{TABLE II in \\ Martin et al.\,\cite{Martin1998}}\\%\hline
        
        2H + H $\rightarrow$ H$_2$ + H & 
        \makecell{TABLE XVII, ln.29 in  \\ Jones et al.\,\cite{Jones1973}}\\\hline
        
        H$_2$ + H$_2$ $\rightarrow$ 2H + H$_2$ & 
        \makecell{TABLE II in \\ Martin et al.\,\cite{Martin1998}}\\%\hline
        
        2H + H$_2$ $\rightarrow$ H$_2$ + H$_2$ & 
        \makecell{TABLE XVII, ln.29 in  \\ Jones et al.\,\cite{Jones1973}}\\\hline
        
        H$_2$ + e$^-$ $\rightarrow$ H + H$^+$ + 2e$^-$ & 
        \makecell{TABLE XII in \\ Yoon et al.\,\cite{Yoon2008}}
    \end{tabular}
    \caption{Reaction channels implemented in HYQUP simulation model}
    \label{tab:reactions}
\end{table}

The previous hydrogen reaction system of the HYQUP model\,\cite{Mewes:2023} is replaced with an improved version, explicitly modeling the particle species H, H$^+$ and now also H$_2$. As before, the total reaction mass source of a particle species $\alpha$ is a sum over all reactions $i$, i.e.
\begin{equation}
    R_\alpha = m_\alpha \sum_i c_{i\alpha} r_i,
    \label{eqn:ReactionSourceTerm}
\end{equation} 
where $m_\alpha$ is the particle mass, $c_{i\alpha}$ is the stoichiometric number (the change of number of particles per reaction $i$) and $r_i$ is the reaction rate. The reaction rates are calculated as 
\begin{align}
    r_i &= k_i \prod_\beta {n_\beta}^{b_{i\beta}} \label{eqn:ReactionRate} 
\end{align}
where $n_\beta$ is the number density of reactant species $\beta$, and $b_{i\beta}$ are the stoichiometric coefficients of the reactants (number of reactants needed per reaction). The reaction constants $k_i$ are independent of the densities, and sourced as summarized in Table\,\ref{tab:reactions}.

% e-collisions
The first pair of reactions is ionization and recombination by electron impact. An analytical model for the 3-body-recombination is recommended by the NRL plasma formulary\,\cite{NRL2019}, accounting also for the recombination into any excited state, assumed to be followed by de-excitation. The total ionization rate, including multi-step processes through collisional excitation, is approximated by applying the principle of detailed balancing\,\cite{boltzmann1964lectures, tolman1979principles} as described in previous work\,\cite{Mewes:2023}. %This choice emphasizes accuracy of the recombination rate, which is slower and well-resolved in the benchmark, over the ionization rate, while retaining a consistent equilibrium state. 

% radiative
The radiative recombination rate recommended by the NRL plasma formulary is also included, but observed negligible under the conditions investigated here. Radiative ionization is completely neglected, as we assume the plasma to be optically thin.
% H collisions (apparently not in Sim. Potential source [Drawin1969])
% At low ionization, late in the cooling of the plasma the electron impact reactions are rare. Here the 3 body collisions between with neutral atoms take a significant role in recombining the remaining plasma.

% H2 things
Four types of dissociation reactions are included for molecular hydrogen. The first, by electron impact, is approximated by the ionization reaction for an atom with the binding energy of the molecule, recommended by the NRL plasma formulary. Two second and third types, by neutral atom or molecule impact, are implemented according to Martin et al.\,\cite{Martin1998}. They are important in the early heat-up of the discharge, where the low ionization prevents electron impact processes.
The fourth type is electron impact ionization of hydrogen molecules, followed immediately by the dissociation of the molecular ion. The rate equation for this is taken from Yoon et al.\,\cite{Yoon2008}. 
The reverse reaction rates for the first type are set by detailed balancing. For the second and third type they are taken from Jones et al.\,\cite{Jones1973}, estimating neutral atom reaction by the molecular rate coefficient. The fourth type is not directly reversible due to its multi-step nature.

% Boundary conditions
\subsection{Boundary Physics}
%\mathis{ref the figure?}

When a plasma interacts with a surface a \emph{plasma sheath} is formed, where quasi-neutrality is breaks down\,\cite{Bohm1949, Godyak1982}. This includes several ways in which the plasma exchanges energy with the surface. Under the conditions investigated here, the sheath at the wall is much thinner than the capillary radius, making it reasonable to approximate the inner sheath boundary as the simulation boundary.

% Ablation of wall material is neglected in the model.
% What is the sheath?
Hot plasma electrons can penetrate into a surface, while the much heavier, slower ions can not. This results in a negative charge building up in the capillary walls that separates nearby plasma charges. The ions entering the sheath are pulled onto the surface and recombine with electrons there.

% Sheath model
The mechanisms governing sheath formation are kinetic and therefore cannot be captured by the HYQUP quasi-neutral fluid model, but simplified analytical models can approximate the influence of a sheath as a boundary conditions. A simple model for electrostatic sheaths on unbiased surfaces was recommended by Pekker et al.\,\cite{Pekker2014}, and has been implemented in HYQUP with minor modifications: 
% Boundary reactions
The original model of Pekker et al. assumes the plasma to be in local thermal equilibrium and the sheath to be collisional, so any ions recombined at the wall are re-ionized on the way back through the sheath to the plasma. 
The HYQUP model does not assume local thermal equilibrium and the sheath is almost collisionless in the discharge regime considered here. The boundary is assumed to recombine ions, while re-ionization is calculated by the reaction model. The ionization energy contribution is therefore subtracted from the electron heat flux to the wall, leaving it as
\begin{equation}
    q_e = - j_{i,\text{wall}} \left( \phi_{\text{float}} + \frac{m_i v_s^2}{2 e} \right) - j_{e,\text{wall}}\frac{2 k_B T_e}{e},
    \label{eq:WallSheatElectronHeatFlux}
\end{equation}
where $j_{e,\text{wall}}$ and $j_{i,\text{wall}}$ are the electron and ion currents towards the wall, $\phi_{\text{float}}$ is the electric potential between wall and plasma, $m_i$ is the mass of the ions, $v_s$ is the sonic velocity of ions, $e$ is the elementary charge, $k_B$ is the Boltzmann constant and $T_e$ is the electron temperature. 

The recombination surface reaction rate at the boundary is
\begin{equation}
    r_{\text{rec, wall}} = \frac{j_{i,\text{wall}}}{e}.
    \label{eq:WallSheatRecombinationRate}
\end{equation}

% Boundary neutral flux
Another change is made to the heavies temperature boundary condition. Instead of applying a Dirichlet condition at the wall temperature, the thermal conductivity of neutrals and the loss of the ions' thermal energy upon contact with the wall are used to estimate the heat flux as
\begin{equation}
    q_h = - j_{i,\text{wall}} \frac{3 k_B}{2 e} (T_h - T_w) - \kappa_n \frac{T_h - T_w}{\Delta x}
    \label{eq:WallSheatHeaviesHeatFlux}
\end{equation}
with the heavies and wall temperatures $T_h$ and $T_w = 300\,\text{K}$, and the width of the sheath $\Delta x$ approximated as an electron Debye radius $\lambda_D = \sqrt{\epsilon_0 k_B T_e/(n_e e^2)}$. Doubling the estimate to $\Delta x = 2\lambda_D$ has negligible effect on the results ($<1\%$).
Physically, the modification of the heavies boundary condition compared to the suggested Dirichlet condition results in similar behavior, but the weaker constraint eases the computational effort.

% What about electrodes?
This sheath model represents the interaction with electrically unbiased surfaces, but is also employed as an approximation for the electrode surfaces here. Most of the plasma bulk is much closer to the wall than the electrodes and dominated by transverse dynamics.

\section{Experimental Benchmark}
\label{sec:bench}
%\begin{itemize}
%    \item Fig: Lab Setup
%    \item ADVANCE lab
%    \item OES setup
%    
%    \item \green{Convoluted Emission model} \mathis{Details in Appendix}
%    \item \green{Fig 6): Simulation valid over broad parameter range} \mathis{mention low current density issues? 2mm in appendix?}
%    \item \green{Fig 7): Some longitudinal resolution is possible. Good qualitative agreement.}
%\end{itemize}

\subsection{Experiment Setup}

\begin{figure}[tb]
    \centering
    \includegraphics[width=\linewidth]{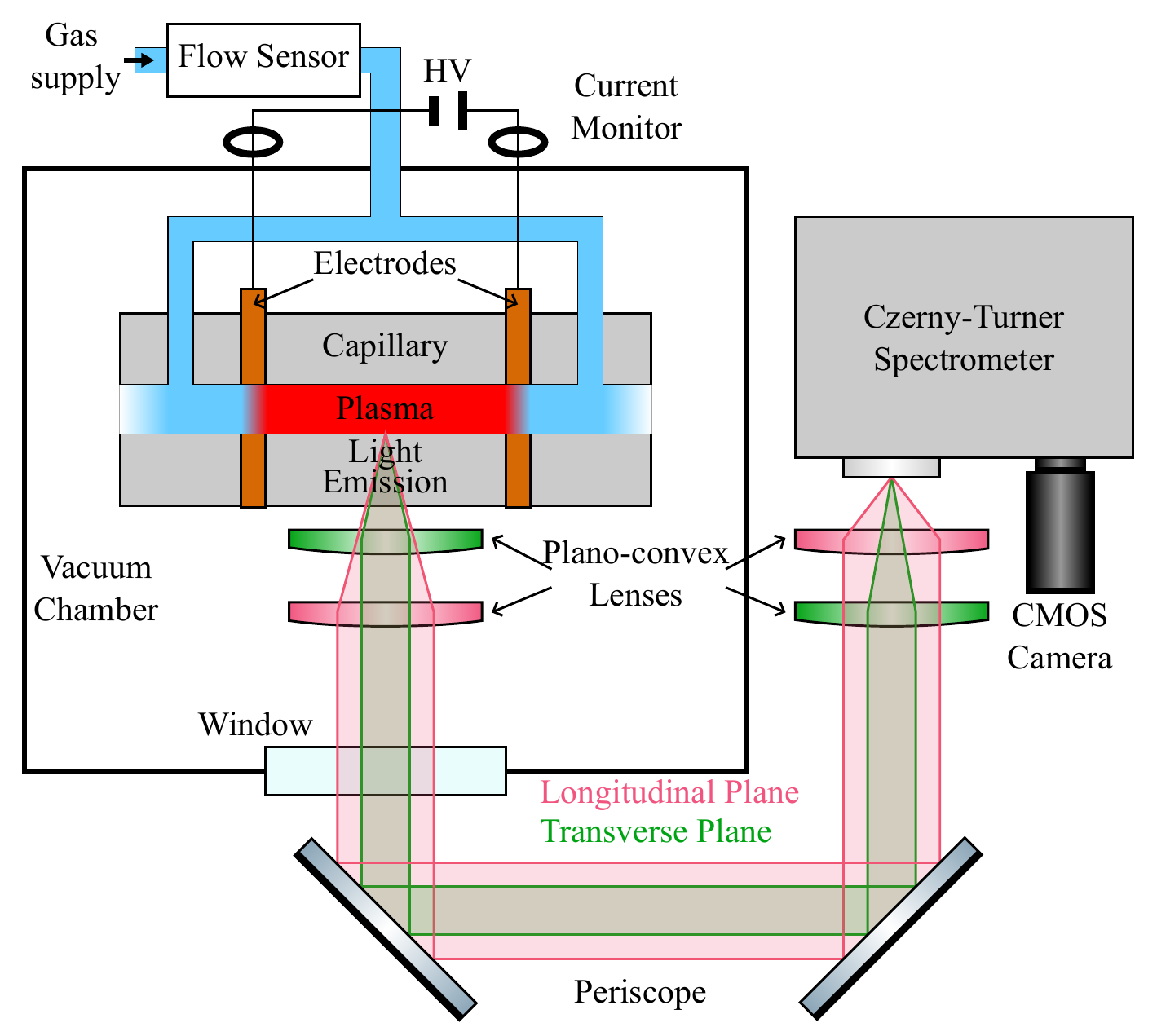}
    \caption{Schematic of the experimental setup: The capillary inside a vacuum chamber is filled with gas and the discharge is triggered. Emitted plasma light is captured by an optics system and analyzed by a spectrometer with a CMOS camera.}
    \label{fig:exp_setup}
\end{figure}

The experimental measurements for this article were performed at the ATHENA Discharge Development and Characterization Experiment (ADVANCE) laboratory at DESY\,\cite{garland:ipac2022-wepopt021, Huck2024-MTmeeting}, using the setup lined out in Fig.\,\ref{fig:exp_setup}.
The 100\,mm long, 1.0\,mm in diameter capillary, with internal plane copper electrodes spaced by 50\,mm, was placed in vacuum and supplied with neutral molecular hydrogen gas at a controlled mass flow rate, varying the pressure inside the capillary in the range 2-12\,mbar.
Plasma was ignited in the section between the electrodes at 2\,Hz repetition rate by 12-27\,kV pulses with approximately 400\,ns duration half width at half maximum (HWHM).
The peak current was 200-600\,A, measured by wide-band current monitors connected to a 2.5 GHz analog-to-digital converter.
% From Pearson electronics

The electron density $n_e$ was inferred using optical emission spectroscopy (OES) over the above ranges of gas pressures and discharge voltages \cite{Gigosos2003, Mijatovic2020, Konjevic2012, Griem1974, Griem1997, Griem2000, Garland:2021}.
The spectrum of radiation emitted by the plasma was observed using a wavelength-calibrated Czerny-Turner-type imaging spectrometer coupled to a gated, intensified CMOS camera. 
Specifically, the width of the Hydrogen-alpha (H$_\alpha$) line was measured, which is broadened by three effects: Stark broadening, which produces a Lorentzian line shape \cite{Gigosos2003}, and Doppler and instrument broadening, both contributing Gaussian components.
The instrument broadening HWHM was measured to be $\Delta\lambda_{\mathrm{inst.}}=0.078\,\mathrm{nm}$.
%\textcolor{red}{Was Doppler broadening accounted for in the analysis? It may not matter for the narrative now we only compared line widths.}
An effective electron density $n_e$ is reconstructed from the measured linewidth by fitting a Voigt function to the data (a convolution of a Gaussian and a Lorentzian distribution).
The HWHM of the Lorentzian component $\Delta\lambda$ can be converted to $n_e$ using
\begin{equation}
    n_e = \left( \frac{\Delta\lambda [\text{nm}]}{1.098} \right)^{1.47135} \times 10^{23} \, \text{m}^{-3},
\end{equation}
following the work of Refs.\,\cite{Gigosos2003} and \cite{Konjevic2012}.

The linewidth was measured as a function of both delay after the initiation of the discharge and longitudinal position.
This was accomplished using two pairs of plano-convex cylindrical lenses, which provided a magnification of 0.17 along the length and 1.0 across the width of the plasma. This setup enabled the full longitudinal line-emission profile to be captured in a single image."

\subsection{Simulation Pipeline}

\begin{figure}[tb]
    \centering
    \includegraphics[width=\columnwidth]{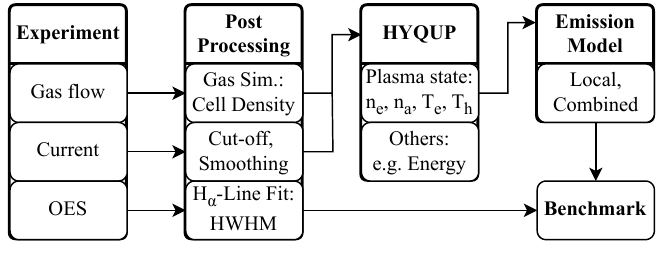}
    \caption{Schematic of the simulation pipeline used for the benchmark.}
    \label{fig:Pipeline}
\end{figure}

Comparing experimental measurements and simulation results requires a well-controlled start-to-end pipeline as depicted in Fig.\,\ref{fig:Pipeline}. The gas flow rate and the current pulses measured in the experiment are processed and put into respective initial conditions and boundary conditions of the plasma simulation. The simulation results are used in an emission model to construct an OES signal analogous to the one observed in the experiment.

% Details on experiment post
Gas flow simulations, using ANSYS, of the gas supply system and the capillary were used to determine the relation between the flow rate from the gas supply and the pressure in the center of the capillary. From these simulations, this relation was observed to be
\begin{equation}
    p\,\mathrm{[Pa]} = 73.79 \times (f_{\text{gas}}\,\mathrm{[ml_n/min]})^{0.57},
    \label{eq:flow_to_dens}
\end{equation}
where $f_{\text{gas}}$ is the volumetric normalized gas flow rate. This was used to set a homogeneous initial gas density in the HYQUP simulations.

To ease the computational effort, the measured current pulse is de-noised via Gaussian convolution, and low currents after 1.5\,µs are cut off. The pulse is used as a constraint in the plasma simulation, fixing the total current flowing through the electrode surfaces in Eq.\,\eqref{eq:electrode_current}.

% Other initial conditions are guessed
Given a lack of knowledge about the exact gas conditions and the electrical breakdown process, we estimate the remaining initial conditions. The initial temperature of both electrons and heavies is assumed to be room temperature everywhere. This has no significant influence on the simulation results, which quickly reaches temperatures on the eV level. The mixture is initialized as homogeneously 0.1\,$\%$ ionized and 1\,$\%$ neutral atomic, as a modest ionization fraction is required for the discharge to further ionize the medium. The remaining 98.9\,$\%$ are molecular.

% The example simulation and geometries
The hydrodynamic simulation of the entire cell geometry in cylindrically symmetric, longitudinally mirrored coordinates, as shown in Fig.\,\ref{fig:cell}, was only performed for one example case, reproducing the experiment with a 20\,kV discharge in 8.6\,mbar gas pressure. This case is used as the example throughout the article.
Results from simulations reduced to the marked simulation box in Fig.\,\ref{fig:cell}, encompassing the discharge region and applying an open boundary for outflowing plasma were found to be indistinguishable from the results of the full simulation, while significantly reducing the computational load. This is hereafter referred to as 2D simulations. A further reduction to a radial slice through the center of the capillary, neglecting longitudinal flow, is hereafter referred to as 1D simulation.

% details on Sim post
% The plasma light emission measured in the experiment are a convoluted quantity. Light is emitted by the entire volume of the capillary, but the resolution is limited, especially in the narrow transverse dimensions. Different light compositions originating from different areas of the capillary are all collected into a single signal. Additionally the optics and the models relating the spectral properties to plasma properties also introduce some uncertainties.

\begin{figure}[tb]
    \centering
    \includegraphics[width=1\linewidth]{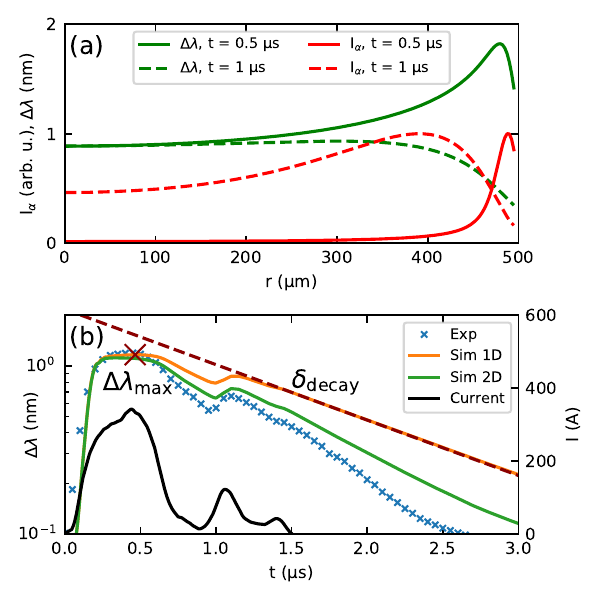}
    \caption{Emission model demonstrated for the example simulation. In (a) the radial distribution of emission intensity and broadened HWHM of the H$_\alpha$ line are shown.
    In (b) the time evolution of the radially averaged spectral line HWHM is shown. The black line shows the measured current profile, used as input in simulations. The $\times$ red cross and red dashed line show the distinctive properties of this curve, namely peak line width $\Delta\lambda_\text{max} = 1.17\,\text{nm}$ and exponential decay rate $\delta_\mathrm{decay} = 0.76\,\text{ps}^{-1}$ (obtained by fitting $\Delta\lambda\propto\exp(-t \delta_\mathrm{decay})$). They are used for benchmarks with experiments in Sec.\,\ref{sec:benchmark}.}
    \label{fig:OES_example}
\end{figure}

% OES model and consequences for experimentalists?
% An emission model is used to construct the $\text{H}_\alpha$ emission line from the plasma simulation results equivalent to the measured line.
It is not known exactly what volume of the narrow transverse dimensions of the capillary the imaging system resolves, but it collects light from a large portion. The $\text{H}_\alpha$ emission spectrum is therefore integrated over the radius of the capillary, and contributions from different plasma densities and temperatures are combined. In order to accurately reproduce the observed spectrum, we append an emission model to the hydrodynamic simulation results.
The details of the emission model are elaborated in appendix\,\,\ref{app:OESmodel}. It calculates the relative emission intensity $I_\alpha$ and the HWHM of the $\text{H}_\alpha$ line $\Delta\lambda$ from the local electron density and temperature. The transverse distributions for two time steps in the example 1D simulation are shown in Fig.\,\ref{fig:OES_example}\,(a). At 0.5 µs, near the peak of the discharge current, the emission originates almost exclusively from the plasma close to the wall, where the line width is large. Later, at 1 µs, the intensity is more homogeneous. These complex dynamics make it difficult to reliably estimate the plasma density near the axis, the region of most interest, with OES.

% The combined spectrum works decently well for this example.
The time evolution of the line width of the radially integrated emission signal is shown in Fig.\,\ref{fig:OES_example}\,(b). 
The 2D simulation, including the longitudinal expulsion of plasma from the capillary, demonstrates much better agreement of the decay process than the 1D simulation, highlighting the importance of the expulsion on longer time scales. The remaining difference to the measurement data, especially at late times, can be attributed to lower accuracy for both simulations and measurements for cool, low-ionization plasma.

The initial kinetic breakdown phenomena, which can not be simulated in this model, lead to a lower reliability of the simulation at early times. An inhomogeneous initial condition can greatly influence the dynamics, because the heating and ionization processes amplify the inhomogeneity. But upon reaching a steady state near the peak of a strong discharge, the results become independent of the initial conditions, making the peak and decay rate of the curve the best targets for a benchmark.

\subsection{Benchmark Results}
\label{sec:benchmark}

\begin{figure}[tb]
    \centering
    \includegraphics[width=1\linewidth]{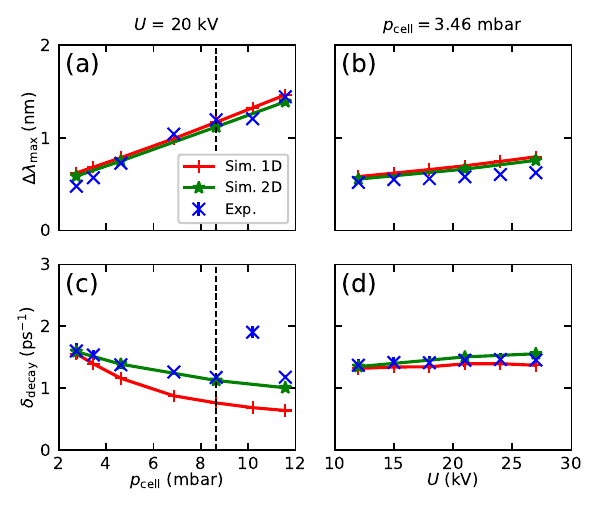}
    \caption{Benchmark summary over broad working parameter space. (a) and (b) show the maximum line width, (c) and (d) the exponential decay rate of each time scan. In (a) and (c) a parameter scan over gas pressure and in (c) and (d) a voltage scan are shown. The vertical line marks the example case used throughout the article.}
    \label{fig:bench_1mm}
\end{figure}

% Parameter bench
For each working point measured in the experiment and simulated, the maximum value and the exponential decay rate are extracted as shown in Fig.\,\ref{fig:OES_example}\,(b) and summarized in Fig.\,\ref{fig:bench_1mm}. Both the trends with the scan parameters, as well as the absolute values, are in good agreement for the full range of gas pressures and voltages, indicating a broad reliability range of the plasma simulation model. The difference between 1D and 2D simulations in Fig.\,\ref{fig:bench_1mm}\,(c) emphasizes the significance of the longitudinal expulsion for the decay curves of the plasma. 

One outlier point in the experimental values in Fig.\,\ref{fig:bench_1mm}\,(c) shows a significantly different time evolution of the signal compared to all other measurements in the scan. It is likely that this measurement originates from a malfunction of the discharge, accidentally operating in an unwanted regime (e.g. partly discharging outside the capillary).

\begin{figure}[tb]
    \centering
    \includegraphics[width=\linewidth]{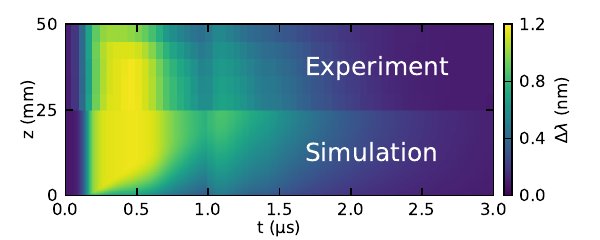} 
    \caption{Longitudinally resolved evolution of line width for the example operating point. The measurement in the top half is compared with the simulation in the bottom half}
    \label{fig:bench_long}
\end{figure}

% Longitudinal benchmark
The experimental measurements also offer longitudinal resolution that is compared to the simulation in Fig.\,\ref{fig:bench_long}. The similar shape of the line width color map confirms that the longitudinal behaviour and expulsion are well represented in the simulation, despite the simplified modelling of the electrode surfaces. %The expulsion dynamics mostly originate from the pressure of the plasma bulk. 

\section{Characterization of Discharge Plasmas}
\label{sec:phys}

% Having established validity of the simulation model with the benchmark, it can be used to gain more understanding of the detailed dynamics, as well as making useful predictions for further development and experimentation.

The HYQUP simulations allow us to explore the full plasma characterization, including local states of the plasma and global parameters such as total energy and power distribution.

\begin{figure}[tb]
    \centering
    \includegraphics[width=1\linewidth]{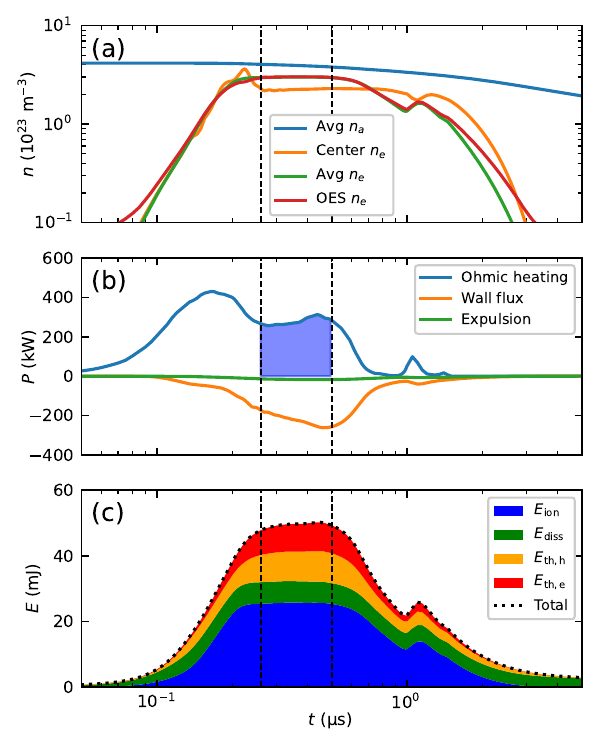}
    \caption{Time evolution of important quantities in the example simulation. In (a) various density values are shown. In (b) the different energy fluxes entering and leaving the plasma are compared. In (c) the internal energy balance of the plasma is shown, with the different energies stacked above each other to sum to the total energy in the plasma. The vertical lines mark the approximate time spent in the steady state.}
    \label{fig:phys}
\end{figure}

% Time evolution
The time evolution of the plasma density is shown in Fig.\,\ref{fig:phys}\,(a) to illustrate the complexities of OES measurements and the importance of start-to-end simulations to shape the plasma profile. The average gas density $n_a$ shows the expulsion of approximately $50\%$ of the original gas content from the discharge region of the capillary (volume between electrodes) within 5 µs. By this time a large fraction of the plasma has recombined and the gas is cooling down, resulting in slowing of the expulsion. 
For accelerator applications, the electron density profile near the capillary axis is a critical parameter. It influences key properties of the wakefield such as accelerating gradient, matching of transverse properties, beam quality preservation, etc.. While OES is a widely used diagnostic method for the electron density, the limited spatial resolution introduces uncertainty, due to the measured emission combining signals from an inhomogeneous plasma volume.

An approximate reconstruction of such a measurement from the simulation, using the OES model described in App.\,\ref{app:OESmodel}, shows differences of up to $50\%$ from the center density. The OES reconstruction is closer, but not identical, to the average density of the plasma. The optical properties of the beamline and capillary wall, as well as the optical properties of the inhomogeneous plasma, contribute to uncertainty in the exact region over which OES signal is integrated. Nevertheless, collecting OES signal from either the whole transverse disk or from a transverse line going through the center yields only a few percent difference in simulated line width, such that the method provides reliable benchmark. It should be noted, however, that extracting the on-axis density only from transverse measurements proves to be challenging.

% Energy Physics?
The dominant energy flux pathways into and out of the plasma are compared in Fig.\,\ref{fig:phys}\,(b). Most of the energy deposited by Ohmic heating is transferred to the capillary wall via thermal flux and recombination reactions, and only a small fraction is lost due to plasma expelled from the capillary. Radiative losses are considered negligible and omitted under the conditions examined here. The total energies, calculated by integrating the energy flux curves, are displayed in Table\,\ref{tab:totalEnergies}.
For the examined working point, approximately 140\,mJ/shot of heat must be removed from the cell to counteract discharge-induced heating during high-repetition-rate operation, not accounting for additional heating from beams or laser pulses. Simple measures could be taken to reduce the wall heat load for this setup. After reaching the steady state condition, which is necessary for shot-to-shot stability, it is maintained for approximately 250\,ns. The energy deposited during this period, marked by the shaded area in Fig.\,\ref{fig:phys}\,(b), could be considered waste. This amounts to 66\,mJ, reducing the required cooling to approximately 74\,mJ/shot.

\renewcommand{\arraystretch}{1.25}
\begin{table}[tb]
    \centering
    \begin{tabular}{c|c}
         &\hspace{5px}E (mJ) \\
        \hline
        Ohmic heating\hspace{5px} & 178.40\\
        Wall flux\hspace{5px} & 141.34\\
        Expulsion\hspace{5px} & 17.24\\
        Remaining\hspace{5px} & 19.81
    \end{tabular}
    \caption{Total energy fluxes of a discharge pulse, obtained by integration of Fig.\,\ref{fig:phys}\,(c)}
    \label{tab:totalEnergies}
\end{table}
%Total Energies in mj
%178.39575124484858
%-141.34201374995294
%-17.242545587176515
%19.811191907719127
%Superfluous energy in mj
%66.18968125605862
Finally Fig.\,\ref{fig:phys}\,(c) shows the internal distribution of energy types throughout the discharge lifetime. Initially electrons are heated via Ohmic heating, increasing the electron thermal energy $E_\mathrm{th,e}$, which is rapidly converted to other forms. At first, the energy is mostly used for the dissociation of hydrogen molecules $E_\mathrm{diss}$, followed by ionization $E_\mathrm{ion}$ after 100 ns. With the growing number of electrons and the decrease of reaction energy drainage, the thermal energies $E_\mathrm{th,h}$ and $E_\mathrm{th,e}$ start to grow and reach their peak shortly after 200\,ns, when full ionization is reached. During the steady state, the ionization energy is approximately equal to all others combined. After 3\,µs, once the plasma has recombined, only the dissociation energy remains significant, as the re-association to molecular hydrogen gas is a slower process.

\section{Summary \& Perspective}
\label{sec:outro}
%\begin{itemize}
%    \item Discharge model is proposed and validated for broad parameter range. 
%    \item OES shown to be complex measure, not representative of wake field background
%    \item Further complexities: Electrode surfaces, low current density, gas species.
%    \item Heat modelling?
%\end{itemize}
To accurately analyze the dynamics of an electrical discharge plasma cell on the nanosecond to microsecond time scales, the quasi-neutral hydrodynamic HYQUP simulation model has been updated to include electrical currents, boundary interactions, an improved reaction model, and magnetization.
Benchmark against experimental data, spanning the parameter space from 2 mbar to 12\,mbar of hydrogen gas pressure and 12\,kV to 27\,kV discharge voltage, established the reliability of the model to investigate a typical parameter range for plasma acceleration. The study also identified limitations in using $\text{H}_\alpha$ emission spectroscopy to infer electron density of the plasma: transverse measurements only give indirect indications on the non-uniform plasma density profile within the capillary. Finally, energy deposition was investigated, and a typical working point for peak density of $\simeq 10^{23}\,\text{m}^{-3}$ deposits on the order of 100\,mJ/shot for a 5-cm discharge plasma. Simple scaling to megahertz operation results in an average thermal load on the order of 10\,kW per centimeter of plasma, although this number could likely be reduced with dedicated engineering effort.

Further, the HYQUP simulation model will be used to study advanced tailoring of plasma sources for plasma-based accelerators, such as active plasma lenses\,\cite{Drobniak:2025}, high-repetition-rate acceleration and injection schemes. Future application to similar plasma devices in other research fields, e.g. plasma guns\,\cite{Winfrey:2012}, is also possible. Expulsion of plasma and gas through in- and outlets warrants further studies as it impacts the entrance and exit ramps of a plasma accelerators stage and the general density profile. Future developments of the method will include models of the anode and cathode plasma sheaths, other gas species, and approximate modelling of the electric breakdown. Finally, a significantly wider capillary would result in much lower current density where the fully ionized steady state is not reached, and further benchmark in this regime would provide a useful complement.

\section*{Acknowledgements}

This work was funded by the Deutsche Forschungsgemeinschaft (DFG, German Research Foundation)—491245950.
This research was supported by the Maxwell computational resources operated at Deutsches Elektronen-Synchrotron DESY, Hamburg, Germany. This work was also supported by PACRI – Grant Agreement n. 101188004. Views and opinions expressed are however those of the author(s) only and do not necessarily reflect those of the European Union or the European Research Executive Agency (REA). Neither the European Union nor the granting authority can be held responsible for them. The authors thank
Maik Dinter,
Sven Karstensen,
Sandra Kottler,
Kai Ludwig,
Amir Rahali,
Vladimir Rybnikov,
Andrej Schleiermacher,
Rachel Wolf,
for experimental support. We thank Advait Kanekar for helpful discussions.

\appendix

\section{Wall Boundary}
\begin{figure}[tb]
    \centering
    \includegraphics[width=\linewidth]{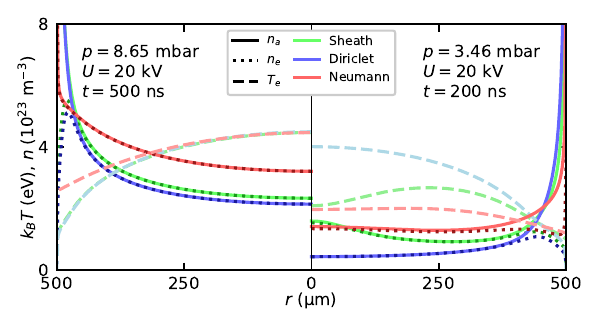}
    \caption{Radial distribution of plasma properties for different boundary conditions in two simulation cases. The style of lines distinguishes plotted properties, while the color stands for the respective boundary conditions.}
    \label{fig:BoundaryComparison}
\end{figure}

In many discharge simulation models the temperature boundary conditions at the capillary walls are simplified to either Dirichlet or Neumann conditions. This can lead to problems demonstrated in Fig.\,\ref{fig:BoundaryComparison}, where the sheath model is compared with a Dirichlet condition $T_e=300\,\text{K}$ and a Neumann condition for thermal flux $q_e = 0$, each using a Dirichlet condition $T_h=300\,\text{K}$ for the heavies.
For a transverse density distribution in the steady state, shown in Fig.\,\ref{fig:BoundaryComparison}(a) the Neumann condition can lead to unphysical results. With heat exchange to the heavies being the only cooling mechanism for the electrons, the thermal energy accumulates and fully ionizes the plasma up to the wall.
The Dirichlet condition works well in the steady state of a strong discharge, showing less than 10\% difference to the sheath model in this example. But at partial ionization in weak discharges or early time steps it significantly changes the dynamics. In Fig.\,\ref{fig:BoundaryComparison}(b) it accumulates more neutral gas at the wall, leading to a much lower plasma density in the bulk, compared with the sheath condition. Here the Neumann condition appears to be a better approximation instead.
Consequently we recommended to use a sheath model to cover both cases well.

\section{Optical Emission Model}
\label{app:OESmodel}
%\begin{itemize}
%    \item Replicate measurement step by step
%    \item \green{Calculate (relative) line emission intensity}
%    \item Calculate line broadening (Stark, thermal, instrument)
%    \item \mathis{Line Shift?}
%    \item Generate local emission profile
%    \item Volumetric integration of weighted emission profiles
%    \item Extracting geometric HWHM
%    \item \green{Fig 5b): Well matching example, longitudinal dynamics (expulsion) important for decay}\red{(electrodes)}
%\end{itemize}

% Intro
The measurements used here were obtained from OES of the Hydrogen Balmer alpha ($\text{H}_\alpha$) emission line. The free electrons cause a Stark broadening of the line-width, making it possible to deduce the local electron density. To make the most reliable comparison for the benchmark the emission spectrum equivalent to the measurement needs to be reconstructed. This requires the calculation of the emission intensity and the line spectrum at each simulation point.

% Emission Intensity
\subsection{Emission Intensity}
The $\text{H}_\alpha$-emissions originate from the decay of the 3rd excited state to the 2nd excited state of a hydrogen atom. The intensity of the emitted light is therefore proportional to the population of the 3rd excited state. This population is used as a relative intensity weight $I_{\alpha}$ when calculating the combined spectrum of a plasma volume. The HYQUP simulation model does not compute excited state populations, it is instead calculated using a model presented by Mitchner\,\&\,Krüger\,\cite{MitchnerKruger1973} for non-equilibrium plasma, that uses a blend of Boltzmann and Saha equilibrium
\begin{equation}
    I_{\alpha} = (1-\zeta) I_{\text{Bol.}} + \zeta I_{\text{Saha}},
\end{equation}
where $I_{\text{Bol.}}$ and $I_{\text{Saha}}$ are the respective equilibrium populations and $\zeta$ is a mixing function depending on the temperature of the plasma. For low temperature the 3rd state is best described in Boltzmann equilibrium, populated by excitation. For high temperature it is in Saha equilibrium with the plasma, populated by electron capture. The switch between the two happens around the temperature where the average electron energy is equal to the ionization energy of the 3rd excited state. Here a smooth mixing function is chosen, defined as
\begin{equation}
    \zeta =  
    \begin{cases}
        1              & \text{for } \epsilon_q^\infty + \frac{\Delta\epsilon}{2} > \epsilon_3^\infty\\
        \frac{1 + \sin\big(
        \pi \frac{k_B T_e - \epsilon_3^\infty}{\Delta\epsilon}\big)}{2}            
        & \text{for } \epsilon_q^\infty - \frac{\Delta\epsilon}{2} > \epsilon_3^\infty\\
        0              & \text{otherwise},
    \end{cases}
\end{equation}
\begin{equation*}
    q = \sqrt{\frac{\epsilon_{\infty}}{k_B T_e}}, \hspace{1cm} \Delta\epsilon = \epsilon_2^4, 
\end{equation*}
\begin{equation*}
    \epsilon_a^b = \epsilon_b - \epsilon_a = \epsilon_\infty * \left(\frac{1}{a^2} - \frac{1}{b^2}\right)
\end{equation*}
where $q$ is an effective state number, marking the boundary between Saha and Boltzmann equilibrium for a given temperature, $\Delta\epsilon$ is the smoothing width, chosen as one state above and below the 3rd, $\epsilon_a^b$ is the energy difference between states $a$ and $b$, and $\epsilon_\infty = 13.6 \text{ eV}$ is the ionization energy of hydrogen. 
The individual equilibrium intensity expressions are defined as 
\begin{align}
    I_{\text{Bol.}} &= n_n g_3 \exp\left(-\frac{\epsilon_1^3}{k_BT_e}\right) \frac{1}{Z_{\text{part.}}(T_e)}, \\
    I_{\text{Saha}} &= n_e^2 \left[ \left(2 \pi m_e \frac{k_B T_e}{h^2}\right)^{\frac{3}{2}} 2 \frac{g_\infty}{g_3} \exp\left(-\frac{\epsilon_3^\infty}{k_B T_e}\right) \right]^{-1}, \\
    Z_{\text{part.}} &= \sum_{a \leq \min(q,10)} a^2 \exp \left[ -\frac{\epsilon_1^\infty (1 - a^{-2})}{k_B T_e} \right],
    \label{eqn:EmissionIntensity}
\end{align}
where $n_n$ and $n_e$ are the neutral atom and electron densities, $g_a$ is the degeneracy of the $a$-th state, and $h$ is the Planck constant. %The partition function $Z_{\text{part.}}$ is calculated from the states in the Boltzmann equilibrium ( $a < q$ ) up to the 10th state at most.
The canonical partition function including all possible bound-states diverges\,\cite{Peierls1979}. The partition function given in Eq.\,\eqref{eqn:EmissionIntensity} does not account for the increasing volume of the hydrogen atom at increasingly excited states, which suppress their contributions\,\cite{Schroeder2020}. Thus, in the partition function $Z_{\text{part.}}$, we choose to include the states according to the Boltzmann equilibrium distribution ( $a < q$ ) up to the 10th state at most.

\subsection{Line Broadening}
% What effects? Stark, thermal, Instrument. Voigt
There are several physical effects leading to broadening of spectral lines. Those considered in the model are Stark broadening by the electric fields of the free plasma electrons, thermal Doppler broadening and a constant instrument broadening of the optics setup.
The local spectral line is modeled as a Voigt function, i.e. a convolution of a Lorentzian and a Gaussian function. The Stark effect specifies the width of the Lorentzian component while the Gaussian is specified by the thermal and instrument broadening. 

% Stark
The Stark broadening model used here was presented by Pardini et al.\,\cite{Pardini2013}, defining the HWHM of the Lorentzian component as 
\begin{align}
    \Delta\lambda_{\text{Stark}} \text{[nm]} &= \frac{1}{20} \frac{1}{a + \frac{b}{\sqrt{n_e [\text{m}^{-3}]}}} \left( \frac{n_e[\text{m}^{-3}]}{c} \right)^{\frac{2}{3}}, \\
    a &= \sqrt{4033.8 - 24.45 (\ln(T\text{[K]}))^2}, \\
    b &= 1.028\times10^{12} + 1.746\times10^8 T\text{[K]}, \\
    c &= 8.02\times10^{18}.
\end{align}

% thermal
The thermal Doppler broadening model is specified for a Gaussian profile\,\cite{Griem1997} with the HWHM 
\begin{equation}
    \Delta\lambda_{\text{th.}} = \lambda_\alpha \sqrt{2 \ln(2) \frac{k_B T_h}{m_H c^2}},
\end{equation}
where $\lambda_\alpha = 656 \text{ nm}$ is the center wavelength of the  emission line and $m_H$ is the mass of a hydrogen atom.

% instrument
Finally the instrument broadening of the experiment setup was found to be $\Delta\lambda_{\text{inst.}} \approx 0.078 \text{ nm}$, using an argon-mercury calibration lamp. The total Gaussian HWHM is then
\begin{equation}
    \Delta\lambda_{\text{Gauss}}^2 = \Delta\lambda_{\text{th.}}^2 + \Delta\lambda_{\text{inst.}}^2.
\end{equation}

% line shift
It should also be noted that the plasma environment induces a shift of the spectral line location. Here the shift is estimated via interpolation on tabulated data from Callaway et al.\,\cite{Callaway1991}.

\subsection{Combined spectrum}
\begin{figure}[tb]
    \centering
    \includegraphics[width=\linewidth]{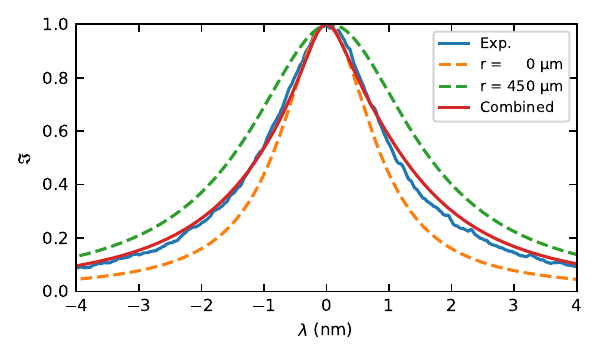}
    \caption{Spectral line example demonstrating the combination of spectral light using the emission model.}
    \label{fig:specline}
\end{figure}

% Pseudo Voigt
The local emission line profile $\mathfrak{I}_{\text{loc.}}(\lambda, \vec{x})$ is constructed as a pseudo-Voigt function on a 10 nm broad spectral grid with a 100 pm resolution, where $\lambda$ is the wavelength relative to the unperturbed emission. Two single point emission lines at the peak of the discharge (t = 500 ns) are calculated from the example simulation and shown in Fig.\,\ref{fig:specline}, demonstrating the large variability with position in the plasma.

% Weighted Integration
An intensity-weighted integral then provides the combined spectrum $\mathfrak{I}_{\text{comb.}}$ of the emission light from a specified plasma volume V, as measured in an experiment. This can be calculated for any slice, or the whole of the capillary
\begin{align}
    \mathfrak{I}_{\text{comb.}}(\lambda) &= \frac{1}{I_{\text{tot.}}} \int_\text{V} \mathfrak{I}_{\text{loc.}}(\lambda,\vec{x}) I_\alpha(\vec{x}) \: d^3x, \\
    I_{\text{tot.}} &= \int_\text{V} I_\alpha(\vec{x}) \: d^3x.
\end{align}
The combined spectrum example in Fig.\,\ref{fig:specline} shows good agreement with the observed measurement.

% Geometric HWHM retrival (Non-voigt! Asymmetry)
Finally the combined HWHM has to be retrieved, but the combined line shape is not a Voigt profile anymore, nor is it symmetric. Here a geometrical method is applied, interpolating the HWHM on both sides from the peak and finally taking the average and quantifying the asymmetry.
\begin{align}
    \Delta\lambda = \frac{\Delta\lambda_{\text{left}} + \Delta\lambda_{\text{right}}}{2}, \\
    \Delta\lambda_{\text{asym.}} = \frac{\Delta\lambda_{\text{left}} - \Delta\lambda_{\text{right}}}{\Delta\lambda}
\end{align}
The asymmetry is generally small ($< 1\%$) for the parameter range examined here.

\bibliography{Bibliography}

\end{document}